\documentclass[useAMS,usenatbib]{mn2e}

\usepackage{amssymb,amsmath}
\usepackage{graphicx}
\usepackage{url}

\newcommand{\be}{\begin{equation}}
\newcommand{\ee}{\end{equation}}

\usepackage{color}

\newcommand{\ion}[2]{{\text{{\sc #1}\,{\sc #2}}}}

\def\aap{A\&A}
\def\apj{ApJ}

\def\apjl{ApJL}
\def\mnras{MNRAS}

\def\prd{Phys.Rev.D}         

\title[Primordial chemistry]
{Non-thermal photons and H$_2$ formation in the early Universe}

\author[C.~M.~Coppola, D.~Galli, F.~Palla, S.~Longo, and J.~Chluba]
{C.~M. Coppola$^{1,2,3}$
\thanks{e-mail: carla.coppola@chimica.uniba.it; galli@arcetri.astro.it;
palla@arcetri.astro.it; savino.longo@ba.imip.cnr.it; jchluba@pha.jhu.edu}, D.~Galli$^3$, F.~Palla$^3$, S. Longo$^{2,3,4}$, 
J.~Chluba$^5$ \\
\\
$^1$Department of Physics and Astronomy, University College London, Gower Street, London WC1E 6BT\\
$^2$Universit\`{a} degli Studi di Bari, Dipartimento di Chimica, Via Orabona 4, I-70126, Bari, Italy\\
$^3$INAF-Osservatorio Astrofisico di Arcetri, Largo E.~Fermi 5, I-50125 Firenze, Italy\\
$^4$IMIP-CNR, Section of Bari, via Amendola 122/D, I-70126 Bari, Italy\\
$^5$Johns Hopkins University, Bloomberg Center 435, 
3400 N. Charles St., Baltimore, MD 21218}

\begin{document}

\date{Accepted 2013 June 5.  Received 2013 June 5; in original form 2013 April 22}


\maketitle

\begin{abstract}
\\
The cosmological recombination of H and He at $z\simeq 10^3$ and the formation of H$_2$ during the dark ages produce a non-thermal photon excess in the Wien tail of the cosmic microwave background (CMB) blackbody spectrum. Here we compute the effect of these photons on the H$^-$ photodetachment and H$_2^+$ photodissociation processes.
We discuss the implications for the chemical evolution of the Universe in the post-recombination epoch, emphasizing how important a detailed account of the full vibrational manifold 
of H$_2$ and H$_2^+$ in the chemical network is. We find that the final abundances
of H$_2$, H$_2^+$, H$_3^+$ and HD are significantly smaller than in previous 
calculations that neglected the effect of non-thermal photons. 
The suppression is mainly caused by extra hydrogen recombination photons and could affect the formation rate of first stars. We provide simple analytical approximations for the relevant rate coefficients and briefly discuss the additional effect of dark matter annihilation on the considered reaction rates.

\end{abstract}


\section{Introduction}

In the era of precision cosmology, the determination of the 
chemical composition of the early Universe requires an accurate evaluation of the
reaction rates of the main chemical processes involved. 
A detailed chemical-kinetic model for the evolution of the homogeneously expanding Universe 
in the post-recombination epoch is also needed to follow the collapse of primordial clouds and hence to study 
the formation of the first-generation stars \citep{Tegmark1997, Abel2002, Bromm2002}. In particular, H$_2$ represents a ``key'' element because of its abundance and coolant properties. For this reason the line emissions associated with molecular hydrogen could in principle give informations about the matter distribution during the phase of pre-reionization of the Universe \citep[e.g.,][]{ciardi, gong}. Over the past years continuous improvements have been made to the modeling
of the early Universe chemistry under non-equilibrium conditions. For example,
\citet[][hereafter C11]{c11}, \cite{lincei} and \cite{c12} demonstrated the importance of taking into account the complete
internal states of chemical partners in a chemical network for the primordial gas,
as well as all non-equilibrium processes occurring at high redshift $z$. 

In this paper we compute the abundance of the main chemical species 
(such as H$_2$, H$_2^+$, H$^-$, HD and H$_3^+$), including the effect of non-thermal photons due to
cosmological recombination of H and He \citep[see][for overview]{Sunyaev2009}, and the radiative cascade 
following the formation of H$_2$. 
The non-thermal photons appear as an excess in the Wien tail of the CMB blackbody spectrum and thus significantly affect the H$^-$ photodetachment and H$_2^+$ photodissociation processes during the dark ages. As we show here, the main effect is caused by the extra \ion{H}{i} recombination photons released at $z\lesssim 100$, limiting the formation of H$_2$, H$_2^+$, H$_3^+$ and HD. 
We also estimate the effect of extra ionizations from annihilating dark matter particles on the H$^-$ photodetachment rate, finding a sensitivity of the early Universe chemistry to this process (see Appendix~\ref{appendixb}).

The paper is organized as follows: in Sections \ref{s1}-\ref{ss1} we
describe the computational methods, providing expressions for the spectral distortion of the CMB introduced by emission processes occurring in past epochs of the expanding Universe. 
The distortion spectra resulting from primordial atomic recombination
and non-equilibrium H$_2$ radiative cascade are then used to evaluate
non-thermal reaction rates for photo-processes involving the main
``catalytic'' species for H$_2$ formation (H$_2^+$ and H$^-$). In
Section \ref{ss1}, we describe the time-dependent kinetics and summarize
the reaction rates and cosmological parameters introduced.
The resulting fractional abundances of several atomic and molecular species adopting the updated rate
coefficients of Coppola et al.~(2011) are discussed
in Section~\ref{results}.

\section{Effect of non-thermal photons}
\label{s1}

Every radiative transition from an upper atomic level $i$ to a lower level $j$ is 
associated with the emission of a photon, causing a spectral distortion $I_{ij}(\nu)$.
Assuming a very narrow emission-profile, the observing frequency $\nu$ at some redshift
$z<z_{\rm em}$, is related to the rest frame frequency, $\nu_{ij}$, of the transition
$i\rightarrow j$ by $\nu=\nu_{ij}(1+z)/(1+z_{\rm em})$. For this reason, the redshift at which the transition
happens is labelled as $z_{\rm em}$. The spectral distortion produced
by the emission process at $z_{\rm em}$ and observed at redshift $z < z_{\rm em}$ can be written as \citep[e.g., see][]{rubino}:
\be
I_{ij}^z(\nu)=\left(\frac{hc}{4\pi}\right)
\frac{\Delta R_{ij}(z_{\rm em})(1+z)^3}{H(z_{\rm em})(1+z_{\rm em})^3}
\label{e2}
\ee
where $H(z)=H_0[\Omega_{\rm r}(1+z)^4+\Omega_{\rm m}(1+z)^3+
\Omega_{\rm k}(1+z)^2+\Omega_\Lambda]^{1/2}$
is the Hubble function and $\Delta R_{ij}$ is related to the level
populations, $N_i$ and $N_j$ of the $i^{\rm th}$ and $j^{\rm th}$ levels by:
\be
\Delta R_{ij} = p_{ij} A_{ij} N_i 
\frac{e^{h\nu_{ij}/k_{\rm B}T_{\rm r}}}{e^{h\nu_{ij}/k_{\rm B}T_{\rm r}}-1}
\left[1-\frac{g_i N_j}{g_j N_i}e^{-h\nu_{ij}/k_{\rm B}T_{\rm r}}\right],
\ee
where $p_{ij}$ is the Sobolev-escape probability, $g_i$ and $g_j$
the degeneracy of upper and lower levels, respectively (both factors are equal
to one in the case of transitions occuring among the vibrational
manifold), $A_{ij}$ is the Einstein coefficient of the transition and
$T_{\rm r}=2.726~(1+z_{\rm em})\,$K \citep{Fixsen1996, Fixsen2009}. 

To evaluate the contribution of spectral distortions to
the reaction rate of a photo-reaction at a given redshift $z$, the integration
over the actual photon distribution should be carried out:
\be
k_{\rm ph}(z)=4 \pi \int_0^\infty \frac{\sigma(\nu)}{h\nu} 
\left[ B_z(\nu)+\sum_{i\rightarrow j} I_{ij}^z(\nu)\right] {\rm d} \nu.
\label{rc}
\ee
Here $\sigma(\nu)$ is the cross section of the photo-reaction as a
function of frequency, $B_z(\nu)$ the Planck distribution at
$T_{\rm r}$ corresponding to the redshift $z$ at which the reaction
rate is calculated, and $I_{ij}^z(\nu)$ the spectral distortion.

\begin{figure}
\begin{center}$
\begin{array}{l}
\includegraphics[width=8.5cm]{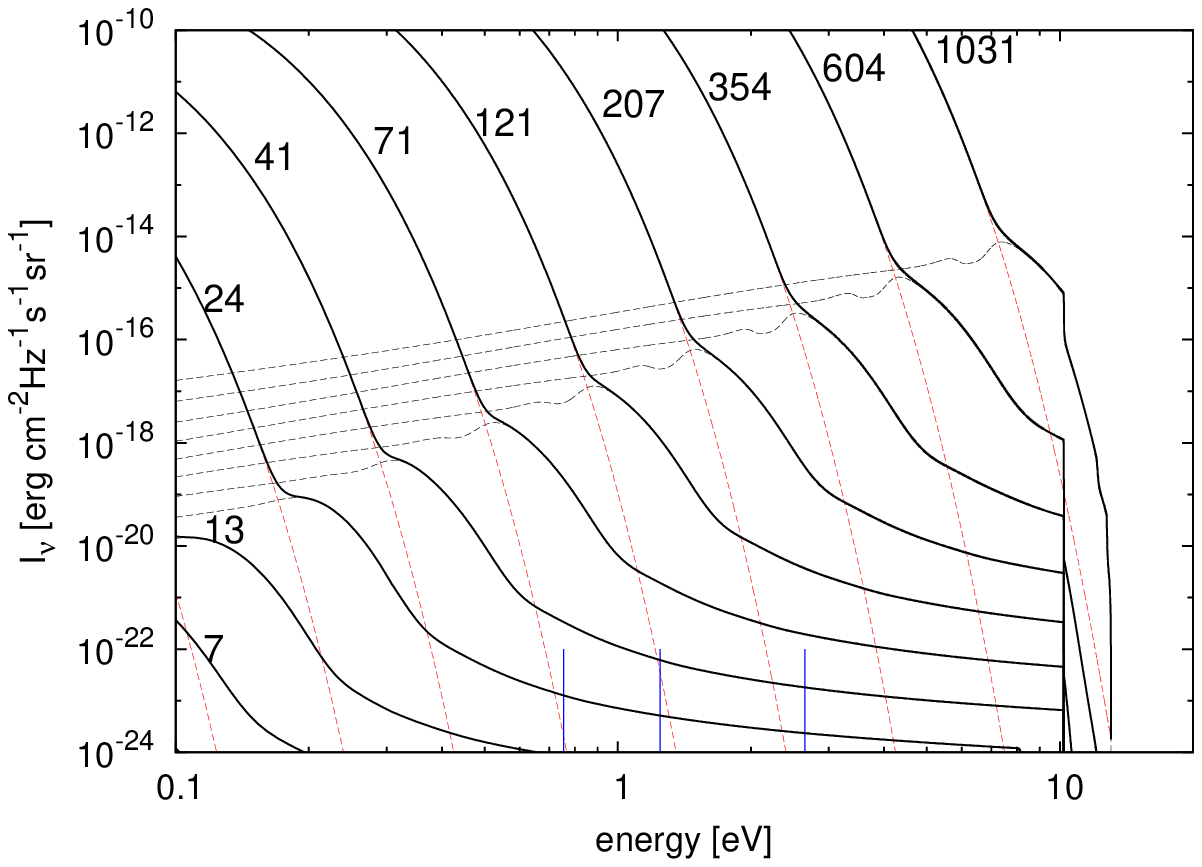}\\
\includegraphics[width=8.5cm]{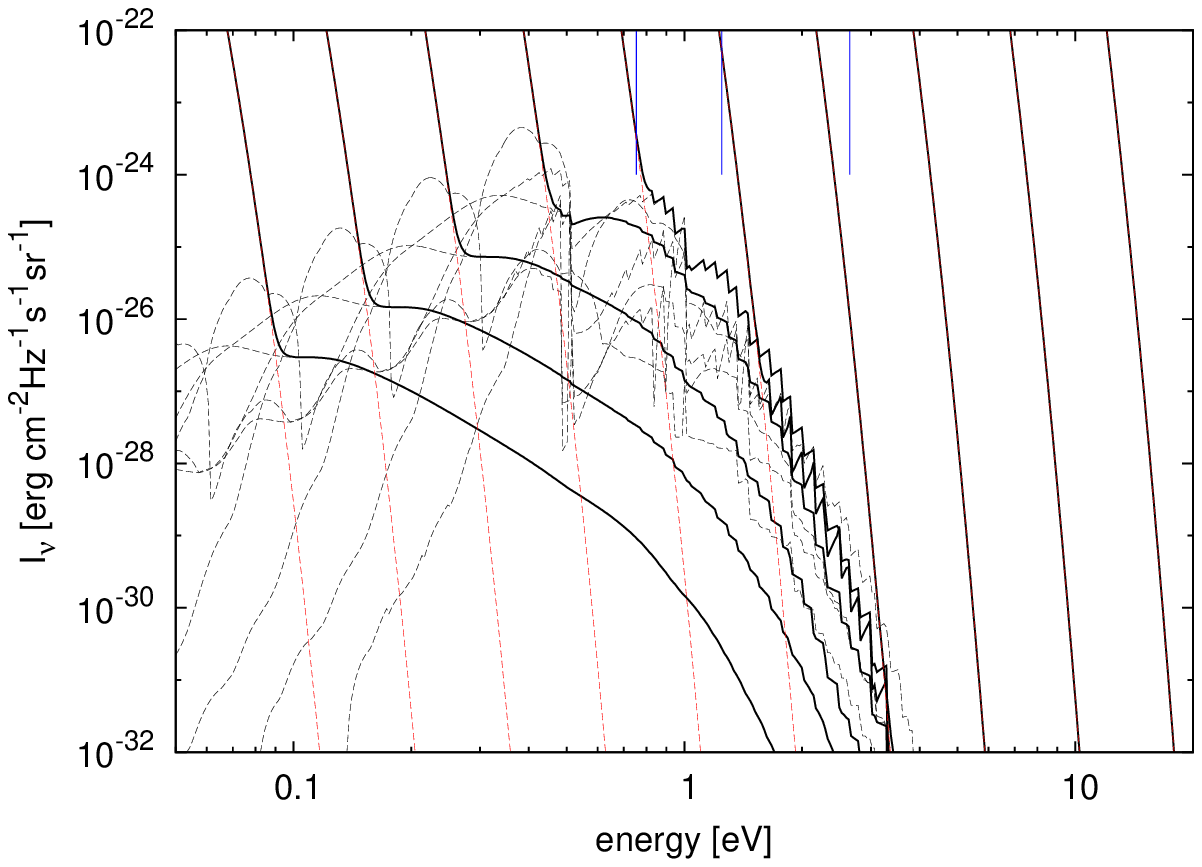}\\
\end{array}$
\end{center}
\caption{Photons spectra at several $z$, corresponding to the values shown in the figure (dashed lines: partial contributions; red and black lines: CMB and distortions, respectively; solid lines: total spectra). {\it Top panel}: blackbody and distortion photons produced by the cosmological recombination of H; {\it bottom panel}: blackbody plus photons produced by the radiative cascade following the non-equilibrium formation of H$_2$ at the same redshifts. The vertical blue lines represent the thresholds for the processes considered in the present work: from the lowest energy, the threshold for H$^-$ photodetachment (0.754 eV), H$^+_2(v=0)$ and H$^+_2(v=6)$ photodissociation (2.65 eV and 1.247 eV, respectively). The value for the highest vibrational level, H$_2^+(18)$, is 0.0029 eV, out of the range of the present figure.} 
\label{f1}
\end{figure}

Several physical and chemical processes can modify the pure blackbody
shape of the CMB \citep[see][for some example related to early energy release]{Chluba2011therm}; in the present calculations, we consider the primordial recombination of H and He and the non-equilibrium radiative cascade of H$_2$ as sources of distortion photons. For the former, the outputs of {\sc CosmoRec}\footnote{\url{www.chluba.de/CosmoRec}} \citep{Chluba2011} are used to evaluate the non-thermal photon contribution. 
For the latter, the values of $A_{ij}$ were calculated
as in C11 averaging over the initial rotational levels and summing
over the final ones the rovibrationally resolved Einstein coefficients
computed by \cite{wolni}; the non-equilibrium level populations calculated in C11 at several $z$ have been used, following the treatment of \cite{c12}. 

To estimate the effect of non-thermal
photons on the H$_2$ chemistry, the main formation channels for molecular
hydrogen, namely the H$_2^+$ and H$^-$ pathways, should be considered
separately.
Figure~\ref{f1} shows the spectrum of the CMB at several
redshifts, including the distortion photons produced by the cosmological 
H recombination and the
H$_2$ radiative cascade. 
For the cosmological recombination radiation, only the emission of the Ly-$\alpha$, Ly-$\beta$ and Ly-$\gamma$ lines and the 2s-1s two-photon continuum are shown, with the computation including detailed radiative transfer effects, such as line feedback and two-photon processes. 
The position of the Ly-$\alpha$ line can be noticed in the upper panel of Fig.~\ref{f1}, while photons above $\simeq 10.2\,{\rm eV}$ are caused by the Ly-$\beta$ and Ly-$\gamma$ transitions. Relative to the Ly-$\alpha$ line these resonances only add a small number of photons in the far Wien tail of the CMB \citep{Chluba2007Lyn} and thus do not affect the final results for the reaction rates at a significant level.
We also omitted the emission caused by transitions among exited states \citep{rubino, Chluba2006fb, b255}, since these only give rise to a tiny derivation relative to the CMB blackbody spectrum.
The reprocessing of helium photons emitted at $z\simeq 2000$ was also taken into account for the computation of the distortion \citep{Chluba2010He}, producing a pre-recombination feature in the \ion{H}{i} Ly-$\alpha$ recombination spectrum \citep[see also][]{Jose2008}.

Helium photons only directly affect the H$_2^+$ and H$^-$ formation rates close to $z\simeq 2400$, so that we did not present their contribution to the CMB spectrum separately.

The lower panel of Fig.~\ref{f1} shows the distortion produced by the radiative cascade following the non-equilibrium formation of H$_2$. At the highest redshift, the largest contribution comes from 
the most energetic $\sim 4.7$ eV transition
between vibrational state $v=14$ and the vibrational ground level of H$_2$ molecules. At lower redshfits, 
the high-$v$ transitions become progressively less important because of
the expansion of the Universe that shifts them to lower energies. The features present in the spectra produced by molecular radiative cascade lines reflect the presence of many transitions with $\Delta v\geq 1$. As a consequence, the spectra are broader than the ones obtained for the atomic recombination (Fig.~\ref{f1}).
As for the cosmological recombination distortion, the photons produced by H$_2$ radiative transitions give rise to an excess in the far Wien tail of the CMB. The extra photons are introduced at late times, during the dark ages, when most of the H$_2$ is forming. In comparison to the recombination radiation it is, however, much smaller and thus only leads to a tiny correction to the reaction rates.

\subsection{H$_2^+$ channel}

Charge transfer between H$_2^+$ and H,
\be
\mathrm{H_2^+}(v)+\mathrm{H}\rightarrow \mathrm{H_2}(v')+\mathrm{H^+},
\label{ce}
\ee
represents the dominant formation channel of H$_2$ at high $z$.
The reaction~\eqref{ce} is exoergic for all vibrational states, $v$, unlike the charge
transfer between $\mathrm{H_2}$ and $\mathrm{H^+}$ that is endoergic
for $v\le 3$, although with low threshold energies (e.g. Krsti\'{c}
\& Schultz~1999, Krsti\'{c} et al.~2002, Krsti\'{c}~2002,2003,2005). For the conditions of the primordial Universe a very efficient collisional way to destroy H$_2^+$ is by dissociative
recombination~\citep{motapon,takagi}. Among the photo-processes, the reaction
\be
\mathrm{H_2^+}(v)+h\nu \rightarrow \mathrm{H}+\mathrm{H^+},
\ee
represents a favourable destruction pathway and it
has been the subject of several theoretical quantum chemistry studies
\citep{dunn, argyros, stancil, lebedev1, lebedev2}. In this work,
we adopt the cross sections calculated by \cite{mihajlov} using a
quantum mechanical method in which the photodissociation process
is assumed to be the result of radiative transitions between the
ground and the first excited adiabatic electronic state of the
molecular ion ${\rm H}_2^+$. These transitions are the results of the interaction
of the electron component of the ion-atom system with the free
electromagnetic field in the dipole approximation. The cross sections
are given as Maxwell-Boltzmann averages (i.e., assuming statistical equilibrium populations for the $J$-substates) of the state-to-state resolved cross
sections $\sigma_{v,J}(\lambda)$ over the rovibrational distribution
function at each temperature:
\be
\begin{split}
\sigma_{\rm ph}(\lambda,T) & =  \frac{1}{Z} \sum_v
\left[\sum_{J~{\rm odd}} \frac{3}{2}(2J+1)
e^{-\frac{E_{vJ}-E_{00}}{k_{\rm B}T}}\sigma_{vJ}(\lambda)\right.\\
& \left.+\sum_{J~{\rm even}} \frac{1}{2}(2J+1)
e^{-\frac{E_{vJ}-E_{00}}{k_{\rm B}T}}\sigma_{v,J}(\lambda)\right ],\\
\end{split}
\ee
where $Z$ is the partition function,
\be
\begin{split}
Z & =\sum_v \left [ \sum_{J~{\rm odd}} \frac{3}{2}(2J+1) 
e^{-\frac{E_{vJ}-E_{0,0}}{k_{\rm B}T}}\right.\\
& \left.+\sum_{J~{\rm even}} \frac{1}{2}(2J+1) 
e^{-\frac{E_{vJ}-E_{00}}{k_{\rm B}T}} \right ].\\
\end{split}
\ee
The temperature used in these equations is the radiation temperature because of the conditions present in the early Universe. As a check in support of this assumption, it can be shown that for the majority of the pairs ($v,v'$), the critical density $n_{cr}=A_{v,v'}/\gamma_{v,v'}$ (i.e. the ratio between radiative and collisional de-excitation coefficients) is much above the mean density of the primordial Universe. Moreover, the hypothesis of thermalization of rovibrational levels has been assumed to make calculations more feasible. This is justified considering the faster relaxation times of rotation with respect to the other molecular degrees of freedom.

\begin{figure}
\includegraphics[width=6cm,angle=-90]{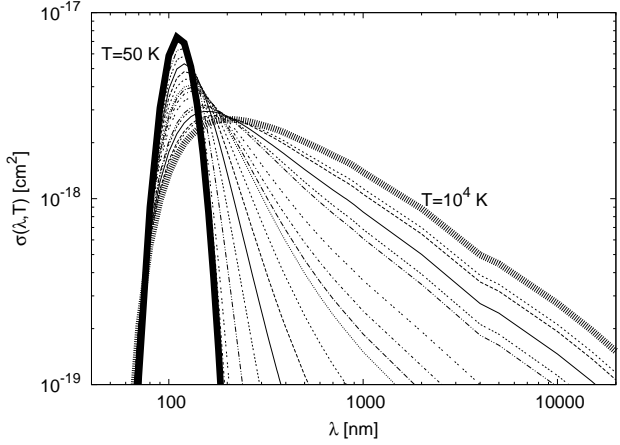}
\caption{H$_2^+$ photodissociation cross sections averaged at several temperatures, from $T=10^4$~K to 50~K. The intermediate curves are for
$T=8400$, 8000, 7000, 6300, 6000, 5000,
4200, 4000, 3500, 3000, 2500, 2000, 1500, 1000,
750, 500, 300, 200, and 100~K.  Calculations have been performed by Mihajlov and Ignjatovi\'{c} using the theoretical method described in Mihajlov et al.~2007 on an extended temperature grid (priv. comm.)}
\label{f2}
\end{figure}

\begin{figure}
\includegraphics[width=9cm]{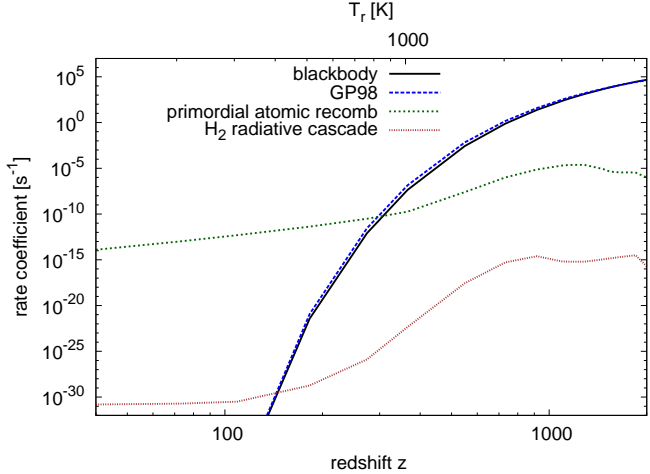}
\caption{H$_2^+$ photodissociation rate coefficient as function of 
redshift $z$ (lower scale) and radiation temperature $T_{\rm rad}$ (upper 
scale).  \emph{Black solid curve}: thermal reaction rate calculated using 
the cross sections shown in Figure~\ref{f1};  \emph{blue dashed curve}: fit by
GP98. \emph{Green dotted
curve}: non-thermal contribution due to cosmological recombination photons computed with 
the cross sections of Figure~\ref{f1}. \emph{Red small dashed curve}: same as above, with photons 
produced by the H$_2$ radiative cascade. It is evident from the figure that distortion photons coming from primordial atomic recombination represent the main non-thermal contribution to the reaction rate.}
\label{f3}
\end{figure}

In Figure~\ref{f2} we show the adopted set of average cross
sections.  Both the CMB blackbody and
non-thermal reaction rates have been calculated using Eq.~\eqref{rc}. The former is compared to the fit given by Galli \& Palla~1998 (hereafter GP98), obtained using
data by \cite{stancil} and \cite{argyros}. The results shown in
Figure~\ref{f3} indicate that the comparison between the thermal reaction rates
is satisfactory at all redshifts. At low temperatures (corresponding to
$z<300$), the dominant effect is due to the non-thermal tails
deriving from atomic recombination. 
The contribution from the H$_2$ cascade in negligible across the temperature interval.

\subsection{H$^-$ channel}

\begin{figure}
\includegraphics[width=8.5cm]{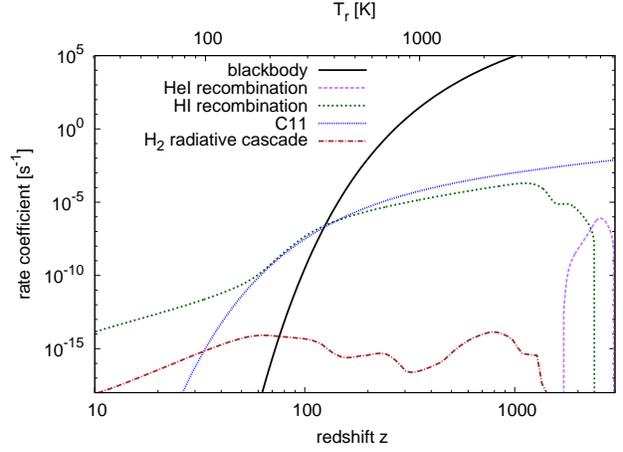}
\caption{H$^-$ photodetachment rate coefficient. Separate contribution of the 
blackbody spectrum, H and He recombination photons, and H$_2$ radiative cascade are shown. As in the case of H$^+_2$ photodissociation, atomic recombination represents the dominant non-thermal contribution.}
\label{f4}
\end{figure}

The formation of H$_2$ at low redshifts is controlled by the associative
detachment process \citep{pagel, cizek, flower, dalgarno, kreckel, schlemmer}:
\be
\mathrm{H^-}+\mathrm{H}\rightarrow \mathrm{H_2}(v)+\mathrm{e^-}.
\label{assdet}
\ee
The reaction responsible for the loss of H$^-$ is the photodetachment process:
\be
\mathrm{H^-}+h\nu \rightarrow \mathrm{H}+\mathrm{e^-}.
\label{phdet}
\ee
For reaction~(\ref{phdet}) we have adopted the analytical fit of
\cite{Tegmark1997} to the cross section data computed by \cite{wishart}.  In
Figure~\ref{f4} we show the H$^-$ photodetachment rate coefficient
obtained considering both the integration over the CMB and
the distortion photons. The former clearly provides the largest contribution at redshifts greater
than $z\sim$100, whereas at lower redshift hydrogen recombination photons contribute significantly. 
The effect of helium photons is restricted to very early times ($z\simeq 2400$) when only very insignificant amounts of chemical elements have formed. The feedback of helium photons on hydrogen also creates extra features in the Ly-$\alpha$ recombination spectrum that leads to non-monotonic behavior of the H$^-$ photodetachment rate at $z\simeq 1800$. It is also important that at high redshifts ($z\simeq 1300$) half of the non-thermal reaction rate is caused by the 2s-1s continuum emission, while in the post-recombination epoch only the \ion{H}{i} Ly-$\alpha$ distortion is important.
As for H$_2^+$ photodissociation, the process of H$_2$ radiative cascade remains negligible at all redshifts.

\section{Reaction rates and kinetics}
\label{ss1}

The chemistry of the early Universe at $z<10^3$ can be described as the kinetics
of a H-He plasma in an expanding medium. For this
reason, the following time-dependent system of ordinary differential
equations \citep[see, e.g.][]{gp98} has to be solved:
\be
\begin{split}
\frac{dN_i}{dt}=&-\sum_l\tilde k_l N_i-\sum_j k_{ij} N_i N_j+\\
&+\sum_n \tilde k_n N_n+\sum_j\sum_m \breve k^{jm}_i N_j N_m, \\
\end{split}
\label{ode}
\ee
where $N_i$ is the abundance of the $i^{\rm th}$ species relative to
the total baryon density, $\tilde k_l$ are the photodestruction rate
coefficients of the $i^{\rm th}$ species via the $l^{\rm th}$ photoprocess; $k_{ij}$ are the destruction rate
coefficients for the $i^{\rm th}$ species for collisions with the
$j^{\rm th}$ chemical partner; $\tilde k_n$ are the formation rate
coefficients of the $i^{\rm th}$ species due to the $n^{\rm th}$ photodestruction
process for the $N_n$ species and $\breve k^{jm}_i$ are the formation
rate coefficients of the $i^{\rm th}$ species due to collisions between
the $j^{\rm th}$ and $m^{\rm th}$ species. Each reaction rate is proportional to the variation of the baryonic density as
a function of time
\be
n_b(z)=\Omega_b\frac{3H_0^2}{8\pi G \mu m_{\rm H}}(1+z)^3,
\ee
where $\Omega_b$ is the baryon fraction, $H_0=100 h$ km s$^{-1}$
Mpc$^{-1}$ is the Hubble constant with $h=0.705$, $G$ is the
gravitational constant, $m_{\rm H}$ is the atomic hydrogen mass,
and $\mu=4/(4-3Y_{\rm p})$ is the mean atomic weight of the gas,
$Y_{\rm p}$ denoting the helium fractional abundance by mass.  The
equations for the radiation and gas temperatures are solved in order
to evaluate the specific velocity of each chemical process in the
kinetics. For the present calculations, the cosmological
parameters from WMAP7 and standard BBN data have been used \citep{newcosmo, iocco}.

For our network, we adopted the rate coefficients summarized in Table~\ref{tabapp}.
The table also provides polynomial fits to the non-thermal contributions to the photodetachment
of H$^-$ and photodissociation of H$_2^+$ rates.
When applicable, the complete vibrational manifolds of H$_2$ and
its cation were used, both in LTE approximation in the entrance
channel and as sum of contributions in the exit channel. 

\section{Results}
\label{results}
\begin{figure}
\begin{center}$
\begin{array}{l}
\includegraphics[width=8.5cm]{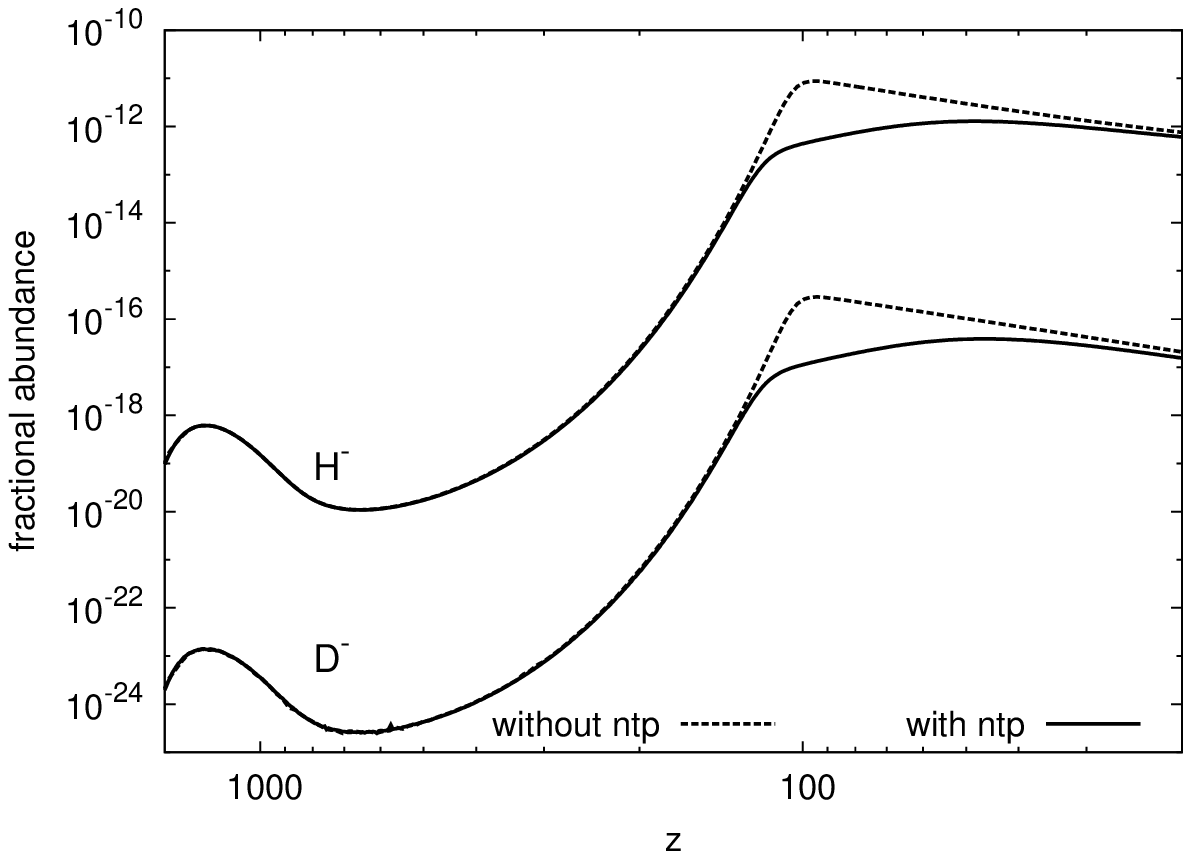} \\
\includegraphics[width=8.5cm]{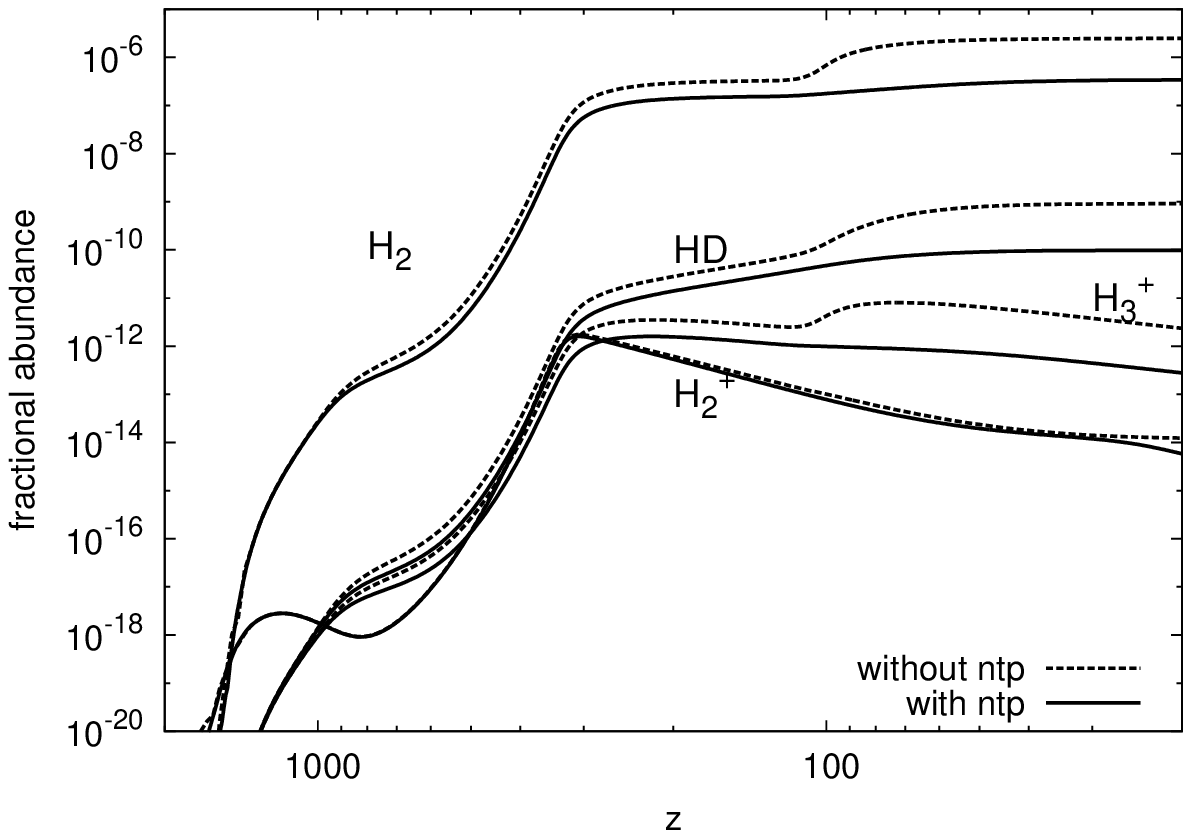} \\
\end{array}$
\end{center}
\caption{Fractional abundances of selected species:
H$_2$, HD and H$_3^+$ ({\em top panel}\/),  H$^-$ and D$^-$
({\it bottom panel}\/), with and without non-thermal photons
({\it green dotted} and {\it blue solid curves}, respectively.)}
\label{f5}
\end{figure}

Using the new rates and the improved rate coefficients reported in C11, we determined
the fractional abundances of several atomic and molecular
species with the kinetic model described in the previous section. Figure~\ref{f5}
shows the evolution of H$_2$, H$_2^+$, H$_3^+$ and HD, along with 
that of H$^-$ and D$^-$. 
The main differences with respect to previous studies can be summarized as follows: starting at
high redshifts, the abundances are affected by 
the enhanced H$_2$ destruction channels (H$_2$/H$^+$
charge transfer, dissociation via H and H$^+$ collisions,
photodestruction), H$^-$ photodetachment and modified H/H$^-$
associative detachment.
The first process results in a reduced fractional abundance of H$_2$
at redshifts $z \sim 1000$ , where it reaches values roughly one
order of magnitude smaller than in previous calculations (e.g. Schleicher et al.~2008). 
Consequently, the fractional abundances of HD and H$_3^+$ are reduced
in the same redshift range. This effect is caused by the inclusion of the entire vibrational manifold,
as also found by \cite{capitelli} for the dissociative attachment process of H$_2$
(see Figures~4-6 of C11).

\noindent At lower redshifts, the abundances are affected by the combined effect of the enhanced
photodetachment of H$^-$ and the  decrease of the efficiency of associative
detachment due to non-thermal photons. This result qualitatively agrees with what was found in the
steady-state model by \cite{hirata}, where however no expression
for the non-thermal rate coefficient was given, and here a more detailed treatment for the recombination spectrum is used. The effect of the contribution
of non-thermal photons to the photodetachment of H$^-$ can be appreciated
in Figure~\ref{f5} (bottom panel) at $z<100$. 
Although the freeze-out value of H$^-$ at low
$z$ remains unchanged, the abundance of H$_2$ is reduced by about 70\%
at the epochs when the H$^-$ channel is dominant.  
Importantly, the new evolution reduces the final rise of the H$_2$ abundance at 
$z\sim$100 that characterized all previous calculations. It is also worth noting that,
despite the huge effect of
non-thermal photons on the photodissociation
of H$_2^+$, its abundance is not significantly
affected. This can be understood considering the relatively high
threshold energy for the photodissociation process compared to the 
photodetachment of H$^-$ and to the mean thermal energy available.
Indeed, the integration over the high frequency part of the distortion spectrum is much more favourable for lower thresholds, as it can be derived from Fig.~\ref{f1}.

\begin{figure}
\begin{center}$
\begin{array}{l}
\includegraphics[width=8.5cm]{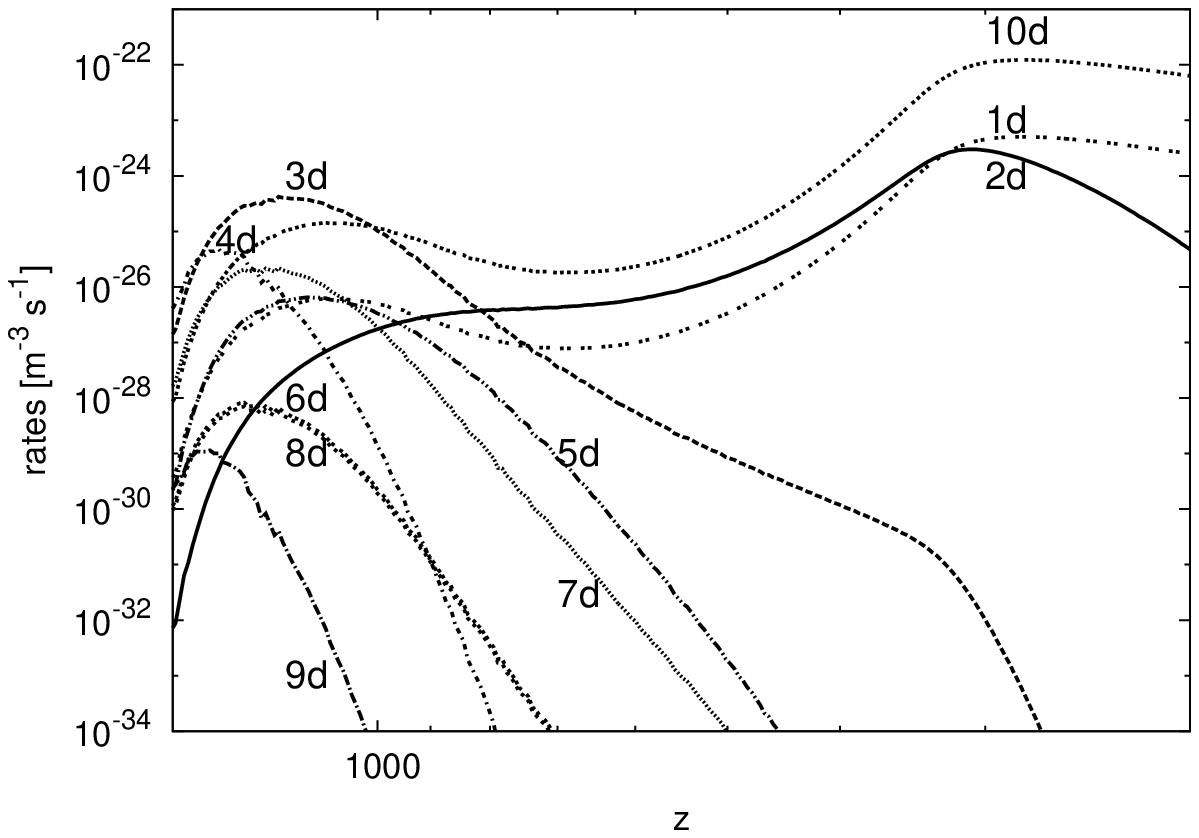} \\
\includegraphics[width=8.5cm]{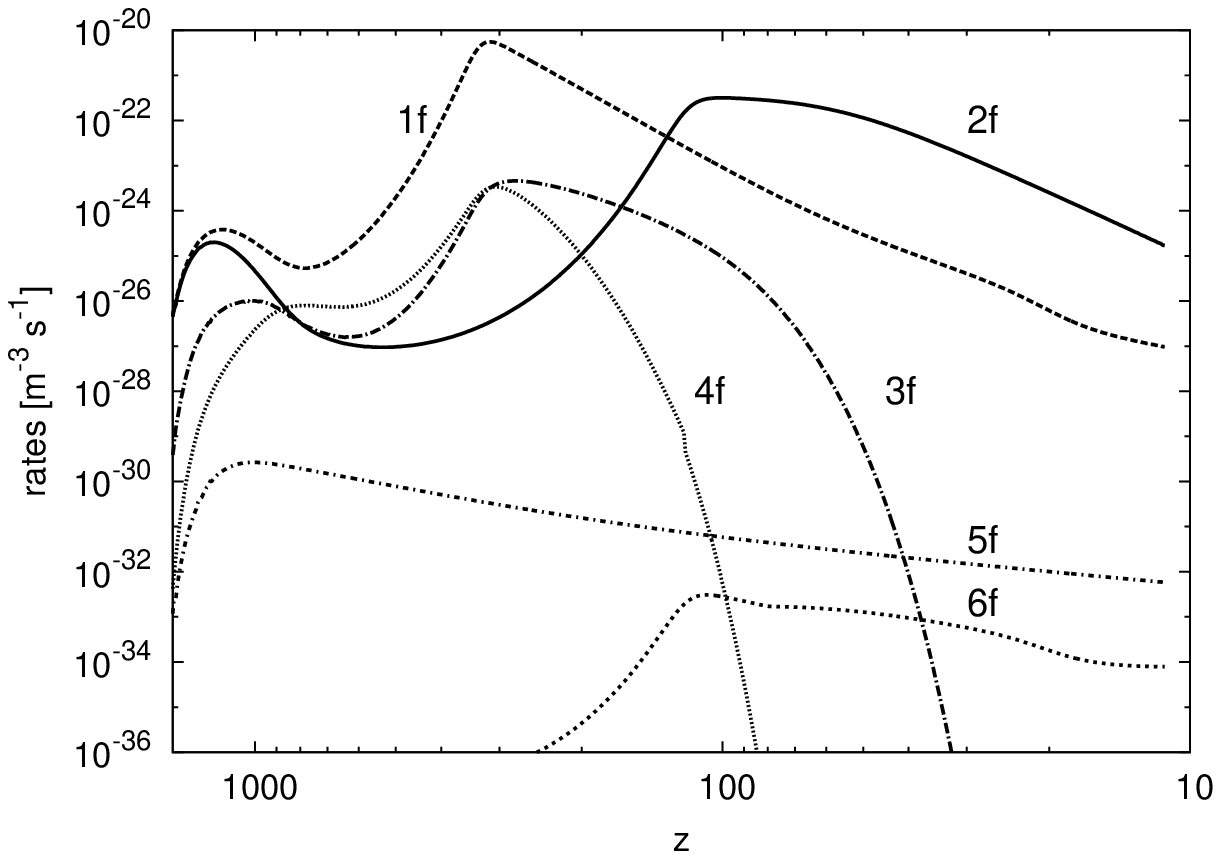} \\
\end{array}$
\end{center}
\caption{H$_2$ destruction ({\it top panel}) and formation ({\it
bottom panel}) rates as a function of redshift $z$. Destruction
processes: ${\rm D}^+ + {\rm H}_2 \rightarrow {\rm HD} + {\rm H}^+$
(1d);  ${\rm D} + {\rm H}_2 \rightarrow {\rm HD} + {\rm H}$ (2d);
${\rm H}^+ + {\rm H}_2 \rightarrow {\rm H}_2^+ + {\rm H}$ (3d);
${\rm H}_2 + h\nu \rightarrow 2 {\rm H}$ (indirect, 4d); ${\rm H}_2
+ {\rm H} \rightarrow 3{\rm H}$ (5d);  ${\rm H}_2 + {\rm H}^+
\rightarrow 2{\rm H}+{\rm H}^+$ (6d); ${\rm H}_2 + e \rightarrow
{\rm H} + {\rm H}^-$ (7d); ${\rm H}_2 + e \rightarrow 2{\rm H} +
e$ (8d); ${\rm  H}_2 + h\nu \rightarrow {\rm H}_2^+ + e$ (9d); ${\rm H}^+ + {\rm H}_2 \rightarrow {\rm H}_3^+ + h\nu$ (10d).
Formation processes: ${\rm H}_2^+ + {\rm H} \rightarrow {\rm H}_2
+ {\rm H}^+$ (1f); ${\rm H}^- + {\rm H} \rightarrow {\rm H}_2 + e$
(2f); ${\rm HD}+{\rm H}^+\rightarrow {\rm D}^+ + {\rm H}_2$ (3f);
${\rm HD} + {\rm H} \rightarrow {\rm D} + {\rm H}_2$ (4f);  $2{\rm
H} + {\rm H} \rightarrow {\rm H}_2 + {\rm H}$ (5f); ${\rm H}_2^+ +
{\rm H}^- \rightarrow {\rm H} + {\rm H}_2$ (6f).}
\label{f6}
\end{figure}

To compare the efficiency of each
process included in the kinetic model, Figure~\ref{f6} shows the destruction and formation rates for H$_2$
as a function of $z$. It should be noted that the
dissociation of H$_2$ via H$^+$ collisions (process labelled ``6d''
in the figure) is one of the most efficient destruction mechanisms,
although it is usually neglected. For the formation channels, the
effect of non-thermal photons is most evident in the channel of 
associative detachment (process``2f'').

\section{Conclusions}

We followed the formation and destruction of the main molecules and
molecular ions in the early Universe, focusing on the effect of non-thermal 
photons produced by the recombination of H and He and 
by the non-equilibrium formation
of H$_2$. We computed the changes in the fractional abundances of
H$_2$, H$^-$, H$_2^+$, H$_3^+$ and on deuterated species such as
D$^-$ and HD. We find that because of high-energy
tails in the photon spectra at several $z$, the efficiency of photodestruction is greatly enhanced,
yielding lower fractional elemental abundances  than in the standard thermal treatment of the chemical kinetics.
 
We also showed that the inclusion of vibrational levels in
the calculation of reaction rates is critical for their
determination at high temperatures when excited
levels are more populated. At high $z$, where these conditons apply, the resulting 
fractional abundances of H$_2$ and H$_2^+$ are reduced by a factor of $\sim$10. 
However, if used in other environments where
molecular hydrogen is more abundant (e.g. during the collapse of primordial clouds),
these new rates are expected to affect more significantly the final molecular abundances.

\section*{Acknowledgments}
The authors acknowledge the referee Steve Lepp for a careful reading of the paper. We are very grateful to Anatolij A. Mihajlov and Lj. M.
Ignjatovic (Insitute of Physics, University of Belgrade) for providing
photodissociation cross section data computed over a wide range of
wavelenghts and temperatures. CMC and SL acknowledge financial
support of MIUR-Universit\`a degli Studi di Bari, (\textquotedblleft
fondi di Ateneo 2010 \textquotedblright) and MIUR-PRIN (grant no. 2010ERFKXL). JC was supported by DoE SC-0008108 and NASA NNX12AE86G. This work has also been
partially supported by the FP7 project ''Phys4Entry'' - grant
agreement n. 242311.

\appendix
\section{Fitting formulae for the rate coefficients}
\label{table}

\begin{figure}
\includegraphics[width=8.5cm]{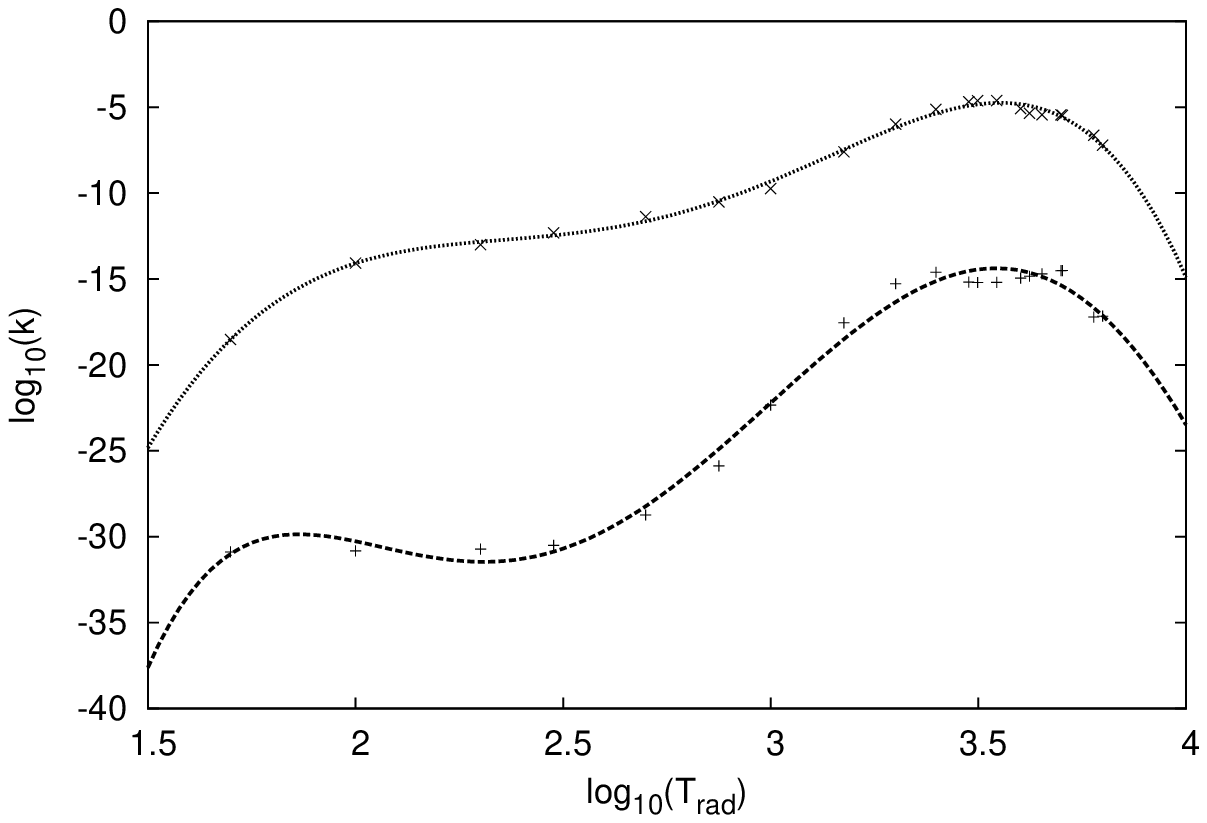}
\caption{Fits ({\it dashed lines}) for the contribution of non-thermal
photons to the photodissociation of H$_2^+$. {\it Upper curve}:
primordial hydrogen recombination contribution; {\it lower curve}:
H$_2$ radiative cascade contribution.}
\label{f7}
\end{figure}

\begin{figure}
\includegraphics[width=8.5cm]{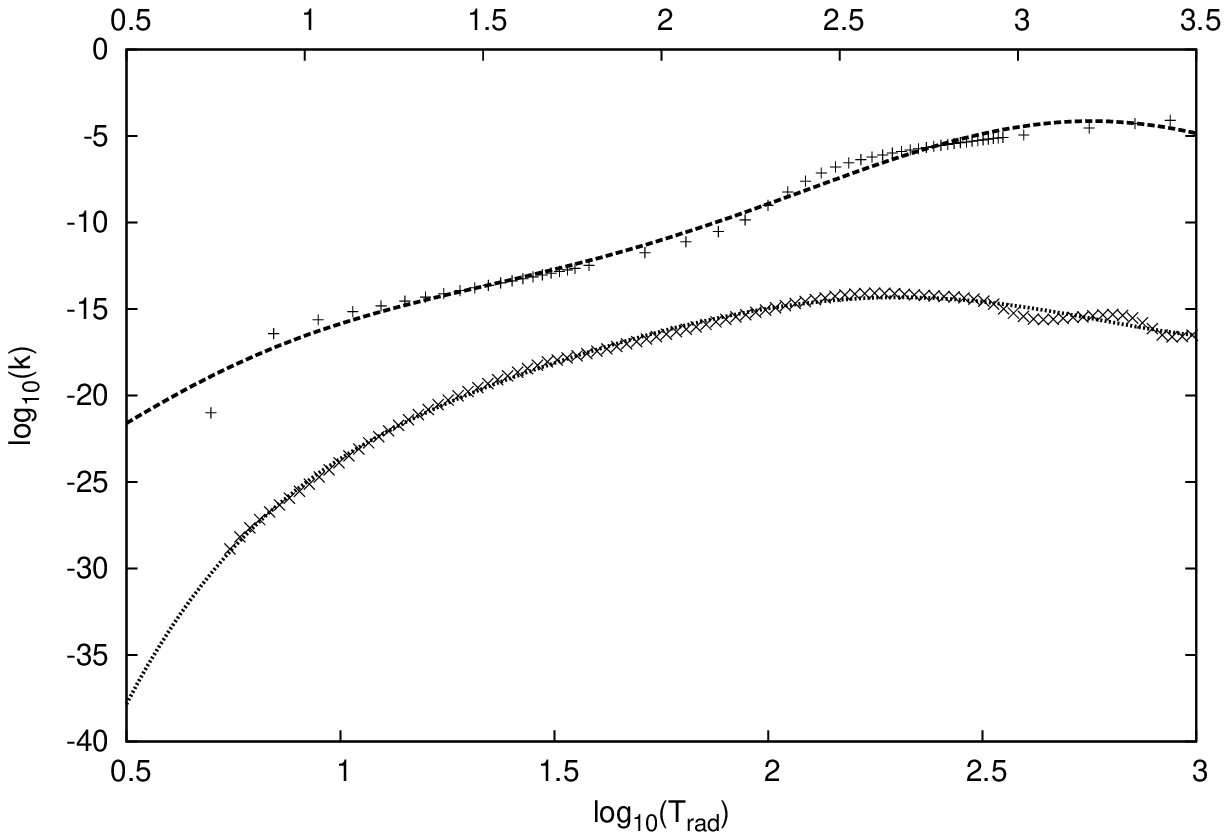}
\caption{Fits ({\it dashed lines}) for the contribution of non-thermal
photons to the photodetachment of H$^-$. {\it Upper curve and
x-axis}: primordial hydrogen recombination contribution; {\it lower
curve and x-axis}: H$_2$ radiative cascade contribution. Their
validity corresponds to the temperature ranges of each axis.}
\label{f8}
\end{figure}

We fitted the reaction rate coefficients for the photodissociation
of H$_2^+$ and the photo detachment of H$^-$ with logarithmic 
polynomials of the form
\begin{equation}
\log k (T_{\rm r}) = \sum_n a_n(\log T_{\rm r})^n.
\label{fit}
\end{equation}
The coefficients of Eq.~(\ref{fit}) are given in Table~\ref{tabapp}, together with the complete set of reaction rates employed in the kinetic model. The temperature of gas and radiation are indicated by $T_{\rm g}$ and $T_{\rm r}$, respectively, and are expressed in K. Natural logarithms are indicated as $\ln$, logarithms to base 10 as $\log$. The results of the fitting formulae are compared to the numerical results in Figures~\ref{f7}-\ref{f8}. 

\section{Dark Matter (DM) annihilation}
\label{appendixb}
Dark matter annihilation or decay leads to extra ionizations of hydrogen and helium atoms in the early Universe \citep{Chen2004, Pad2005}, delaying the cosmological recombination process \citep{Peebles2000} and causing emission of extra recombination photons \citep{dm}.
The additional injection of energy and photons should be taken into account when considering the physical phenomena occurring in the primeval plasma as well as the chemistry. Here we evaluate the effect of extra photons produced by the direct reprocessing of annihilation energy by hydrogen on the rate coefficient of H$^-$ photodetachment. Details on the equations employed and their derivation can be found in \cite{dm}.

\begin{figure}
\includegraphics[width=8.5cm]{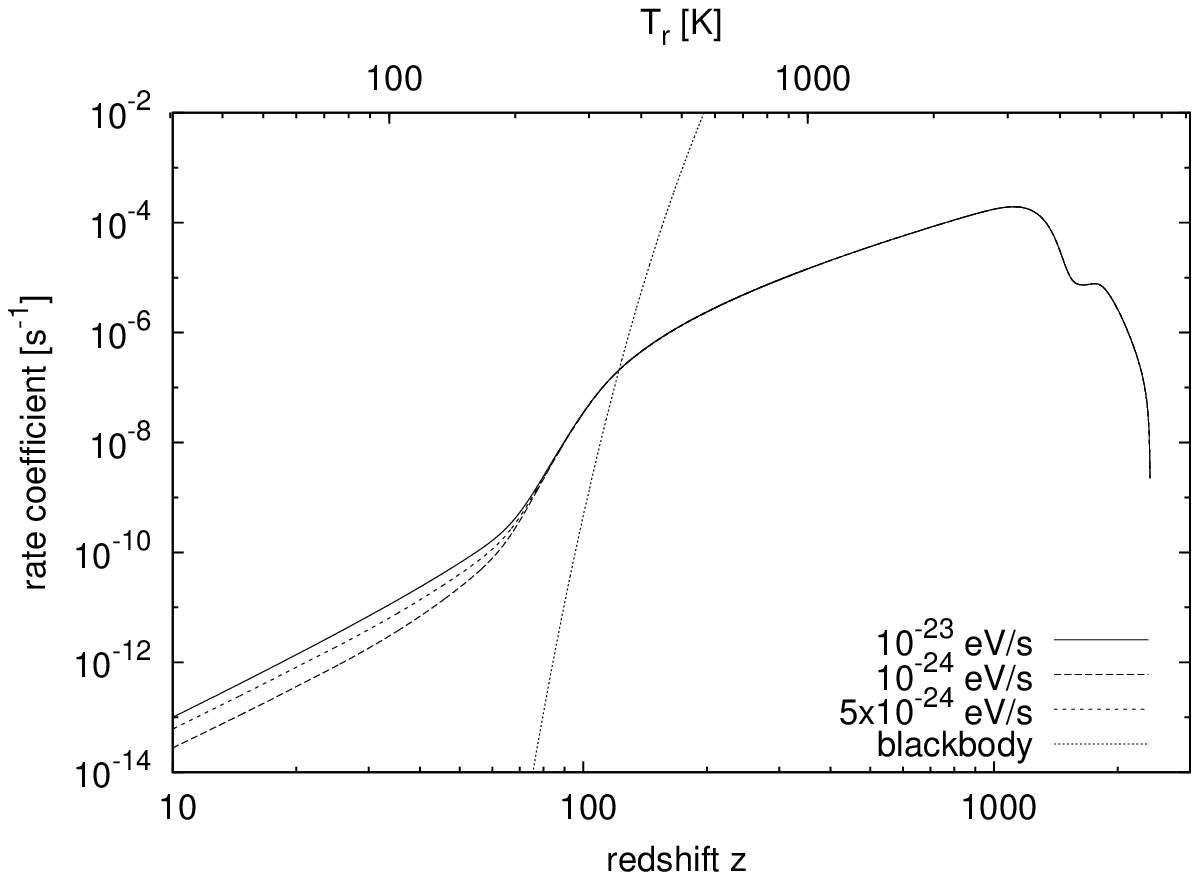}
\caption{H$^-$ photodetachment: effect of dark matter annihilation on the reaction rate. Calculations are reported for different annihilating efficiency, namely $\epsilon_0=10^{-23}$ eV/s, $10^{-24}$ eV/s, $5\times 10^{-24}$~eV/s.}
\label{f9}
\end{figure}

The energy release associated with the annihilation of some DM particle $\chi$ with its antiparticle $\bar{\chi}$ depends on the mass of the particles involved, on the fractional abundances of particle/antiparticle and on the thermally averaged cross section $\langle \sigma v \rangle$ for that process:
\begin{equation}
\frac{\textrm{d}E}{\textrm{d}t}\sim M_{\chi}c^2 \langle \sigma v \rangle N_{\chi}N_{\bar{\chi}}~~[\textrm{eV}~\textrm{cm}^{-3}~\textrm{s}^{-1}]
\end{equation}
Here, results are given as a function of the annihilating efficiency of the particle/antiparticle pair $\epsilon_0$:
\begin{equation}
\frac{\textrm{d}E_\textrm{d}}{\textrm{d}t}=\epsilon_0 \,N_{\rm H}(z)\, (1+z)^3
\end{equation}
In Figure~\ref{f9} we show the cases $\epsilon_0=10^{-23}$, $10^{-24}$ and $5\times10^{-24}$ eV~s$^{-1}$. The presence of extra photons increases the H$^-$ photodetachment rate considerably: the larger the annihilating efficiency the stronger the destruction process. In particular, in the range of annihilating efficiencies shown in the figure and for $z<70$, the enhancement to the rate coefficient scales as:
\begin{equation}
f(\epsilon_0) \sim 1 + 0.44 (\epsilon_0/10^{-24})
\end{equation}
This implies that the early Universe chemistry is not only sensitive to direct ionizations induced by the annihilation products, but also to the reprocessed energy causing additional ionizations of abundant neutral hydrogen atoms and reemission of Ly-$\alpha$ photons.

\begin{table*}
\centering
\begin{minipage}{140mm}
\caption{Reaction rates.}
\begin{tabular}{llrr}
\hline
Process     &   Reaction rates [MKS]         & Ref%
\\
\hline
1) $\mathrm{H}+\mathrm{e^-} \longrightarrow \mathrm{H}^-+h\nu$ & $\footnotesize{1.4\times10^{-24}T_{\rm g}^{0.928}e^{-T_{\rm g}/16200}}$& GP98\\
2) $\mathrm{H}^-+\mathrm{e^-} \longrightarrow \mathrm{H}+2\mathrm{e^-}$ & fit from reference&AAZN97\\
3) $\mathrm{H}^-+\mathrm{H} \longrightarrow 2\mathrm{H}+\mathrm{e^-}$ & fit from reference&AAZN97\\
4) $\mathrm{H}^-+\mathrm{H^+} \longrightarrow 2\mathrm{H}$ & $1.40\times 10^{-13}(T_{\rm g}/300)^{-0.487}e^{T_{\rm g}/29300}$ &LSD02\\
5) $\mathrm{H}^-+h\nu \longrightarrow \mathrm{H}+\mathrm{e^-}$ & &\\
thermal&$0.11{T_{\rm r}}^{2.13}e^{-8823/T_{\rm r}}$ & GP98 \\
non-thermal: atom. recombin. & $\log k=\sum_{n=0}^6 a_n (\log T_{\rm r})^n$  & this work\\
&$a_0=-26.6463$&\\
&$a_1=3.35998$&\\
&$a_2=25.729$&\\
&$a_3=-31.6442$&\\
&$a_4=15.9545$&\\
&$a_5=-3.60013$&\\
&$a_6=0.298272$&\\
non-thermal: H$_2$ radiative cascade&$\log k=\sum_{n=0}^5 b_n(\mathrm{log} T_{\mathrm r})^n$ &this work\\
&$b_0=-81.12$&\\
&$b_1=139.379$&\\
&$b_2=-137.531$&\\
&$b_3=73.0553$&\\
&$b_4=-19.4282$&\\
&$b_5=1.99768$&\\
6) $\mathrm{D}^-+h\nu \longrightarrow \mathrm{D}+\mathrm{e^-}$ & as fit for reaction (5) & S08 \\
7) $\mathrm{HD}^++h\nu \longrightarrow \mathrm{D}+\mathrm{H}^+$ & $(1/2)\times 1.63\times 10^7e^{-32400/{T_{\rm r}}}$ & S08 \\
8) $\mathrm{HD}^++h\nu \longrightarrow \mathrm{D}^++\mathrm{H}$ & $(1/2)\times 1.63\times 10^7e^{-32400/{T_{\rm r}}}$ &S08 \\
9) $\mathrm{HD}^++h\nu \longrightarrow \mathrm{H}^++\mathrm{D}^++\mathrm{e^-}$ & $9.0\times 10^1{T_{\rm r}}^{1.48}e^{-335000/{T_{\rm r}}}$ &S08\\
10) $\mathrm{HD}+h\nu \longrightarrow \mathrm{HD}^++\mathrm{e^-}$ & $2.9\times 10^2{T_{\rm r}}^{1.56}e^{-178500{T_{\rm r}}} $ & S08 \\
11) $\mathrm{D}+\mathrm{H}^+ \longrightarrow \mathrm{D}^++\mathrm{H}$ & $2.0\times 10^{-16}T_{\rm g}^{0.402}e^{-37.1/{T_{\rm g}}}-3.31\times 10^{-23}{T_{\rm g}}^{1.48}$ &SA02\\
12) $\mathrm{D}^++\mathrm{H} \longrightarrow \mathrm{D}+\mathrm{H}^+$ & $2.06\times 10^{-16}{T_{\rm g}}^{0.396}e^{-33.0/{T_{\rm g}}}+2.03\times 10^{-15}{T_{\rm g}}^{-0.332}$ & SA02 \\
13) $\mathrm{D}+\mathrm{H} \longrightarrow \mathrm{HD}+h\nu$ &  $10^{-32}[2.259-0.6({T_{\rm g}}/10^3)^{0.5}+0.101(T_{\rm g}/10^3)^{-1.5}$ &  \\
 &  $-0.01535(T_{\rm g}/10^{3})^{-2}+5.3\times 10^{-5}(T_{\rm g}/10^{3})]^{-3}$ & DI08 \\
14) $\mathrm{HD}^++\mathrm{H} \longrightarrow \mathrm{HD}+\mathrm{H}^+$ & $6.4\times 10^{-16}$ & SLD98 \\
15) $\mathrm{D}+\mathrm{H}^+ \longrightarrow \mathrm{HD}^++h\nu$ & $\log k/10^{-6}=-19.38-1.523\log T_{\rm g}+1.118(\log T_{\rm g})^2$ & \\
 & $-0.1269(\log T_{\rm g})^3$ & GP98 \\
16) $\mathrm{D}^++\mathrm{H} \longrightarrow \mathrm{HD}^++h\nu$ & as fit for reaction (15) & GP98 \\
17) $\mathrm{HD}^++\mathrm{e}^- \longrightarrow \mathrm{D}+\mathrm{H}$ & $7.2\times 10^{-14}T_{\rm g}^{-0.5}$ & SLD98 \\
18) $\mathrm{D}+\mathrm{e}^- \longrightarrow \mathrm{D}^-+h\nu$ & $3.0\times 10^{-22}(T_{\rm g}/300)^{0.95}e^{-{T_{\rm g}}/9320}$ & SLD98 \\
19) $\mathrm{D}^++\mathrm{D}^- \longrightarrow 2\mathrm{D}$ & $1.96\times 10^{-13}(T_{\rm g}/300)^{-0.487}e^{T_{\rm g}/29300}$ & LSD02\\
20) $\mathrm{H}^++\mathrm{D}^- \longrightarrow \mathrm{D}+\mathrm{H}$ & $1.61\times 10^{-13}(T_{\rm g}/300)^{-0.487}e^{T_{\rm g}/29300}$ & LSD02\\
21) $\mathrm{H}^-+\mathrm{D} \longrightarrow \mathrm{H}+\mathrm{D}^-$   & $6.4\times 10^{-15}(T_{\rm g}/300)^{0.41}$ & SLD98 \\
22) $\mathrm{D}^-+\mathrm{H} \longrightarrow \mathrm{D}+\mathrm{H}^-$   & $6.4\times 10^{-15}(T_{\rm g}/300)^{0.41}$ & SLD98 \\
23) $\mathrm{D}^-+\mathrm{H} \longrightarrow \mathrm{HD}+\mathrm{e}^-$ & $1.5\times 10^{-15}(T_{\rm g}/300)^{-0.1}$ & SLD98 \\
24) $\mathrm{D}+\mathrm{H}^- \longrightarrow \mathrm{HD}+\mathrm{e}^-$ & as fit for reaction (22) & S08 \\
25) $\mathrm{H}^-+\mathrm{D}^+ \longrightarrow \mathrm{D}+\mathrm{H}$ & $1.61\times 10^{-13}(T_{\rm g}/300)^{-0.487}e^{T_{\rm g}/29300}$ & LSD02 \\
26) $\mathrm{He}+\mathrm{H}^+ \longrightarrow \mathrm{He}^++\mathrm{H}$ & $4.0\times 10^{-43}T_{\rm g}^{4.74}$ for 
$T_{\rm g} >10^4$ &  \\
 &$1.26\times 10^{-15}T_{\rm g}^{-0.75}e^{-127500/T_{\rm g}}$ for $T_{\rm g}<10^4$ & S08 \\
27) $\mathrm{He}^++\mathrm{H} \longrightarrow \mathrm{He}+\mathrm{H}^+$ & $1.25\times 10^{-21}(T_{\rm g}/300)^{0.25}$ & ZDKL89 \\
28) $\mathrm{He}+\mathrm{H}^+ \longrightarrow \mathrm{HeH}^++h\nu$ & $8.0\times 10^{-26}(T_{\rm g}/300)^{-0.24}e^{-T_{\rm g}/4000}$ & SLD98 \\
29) $\mathrm{He}+\mathrm{H}^+ +h\nu\longrightarrow \mathrm{HeH}^++h\nu$ & $3.2\times 10^{-26}T_{\rm g}^{1.8}
e^{-T_{\rm g}/4000}(1+2\times 10^{-4}T_{\rm r}^{1.1})(1+0.1 T_{\rm g}^{2.04})^{-1}$ & JSK95, ZSD98 \\
30) $\mathrm{He}^++\mathrm{H}\longrightarrow \mathrm{HeH}^++h\nu$ & $4.16\times 10^{-22}T_{\rm g}^{-0.37}e^{-T_{\rm g}/87600}$ & SLD98 \\
31) $\mathrm{He}^++\mathrm{e}^-\longrightarrow \mathrm{He}+\mathrm{H}$ & $3.0\times 10^{-14}(T_{\rm g}/300)^{-0.47}$ & SLD98 \\
32) $\mathrm{HeH}^++h\nu\longrightarrow \mathrm{He}+\mathrm{H}^+$ & $2.20\times 10^2 e^{-22740/T_{\rm r}}$ & JSK95 \\
33) $\mathrm{HeH}^++h\nu\longrightarrow \mathrm{He}^++\mathrm{H}$ & $7.8\times 10^3 T_{\rm r}^{1.2} e^{-240000/T_{\rm r}}$ & GP98 \\
\hline
\end{tabular}
\end{minipage}
\label{tabapp}
\end{table*}

\begin{table*}
 \centering
 \begin{minipage}{140mm}
  \caption{Reaction rates.}
  \begin{tabular}{llrr}
  \hline
   Process     &   Reaction rates [MKS]         & Ref%
  \\
 \hline
34) $\mathrm{H}^-+\mathrm{H}\longrightarrow \mathrm{H_2}+\mathrm{e}^-$ &$\log k=-14.4-0.15(\log T_{\rm g})^2-7.9\times 10^{-3}(\log T_{\rm g})^4$ & C11 \\
35) $\mathrm{H}^++\mathrm{H}\longrightarrow \mathrm{H_2^+}+h\nu$ & $\log (k/10^{-6})=-19.38-1.523\log T_{\rm g}+1.118(\log T_{\rm g})^2 $ &  \\
 & $-0.1269(\log T_{\rm g})^3$ & GP98 \\
36) $\mathrm{H_2}^++\mathrm{H}\longrightarrow \mathrm{H_2}+\mathrm{H}^+$ & $6.4\times10^{-16}$ & GP98 \\
37) $2\mathrm{H}+\mathrm{H}\longrightarrow \mathrm{H_2}+\mathrm{H}$ & $5.5\times 10^{-35}T_{\rm g}^{-1}$ & PSS83 \\
38) $\mathrm{H_2}+\mathrm{H}^+\longrightarrow \mathrm{H_2^+}+\mathrm{H}$&$\ln k=a_0+a_1T_{\rm g}+a_2T_{\rm g}^{-1}+a_3 T_{\rm g}^2$ & C11 \\
  & $a_0=-33.081$\\
  & $a_1=6.3173\times 10^{-5}$\\
  & $a_2=-2.3478\times 10^4$\\
  & $a_3=-1.8691\times 10^{-9}$\\ 
39) $\mathrm{H_2}+\mathrm{e}^-\longrightarrow 2\mathrm{H}+\mathrm{e}^-$ &$1.91\times 10^{-15}T_{\rm g}^{0.136}e^{-53407.1/T_{\rm g}}$ & TT02 \\
40) $\mathrm{H}^-+\mathrm{H}^+\longrightarrow \mathrm{H_2^+}+\mathrm{e}^-$ & $6.9\times10^{-15}T_{\rm g}^{-0.35}$ for $T_{\rm g}< 8000$ & GP98 \\
& $9.6\times10^{-13}T_{\rm g}^{-0.9}$ for $T_{\rm g}>8000$ & \\
41) $\mathrm{H_2}^++\mathrm{e}^-\longrightarrow 2\mathrm{H}$ & $k=\sum_{n=0}^5 a_n T_{\rm g}^n$ & C11 \\
  & $a_0=4.2278\times 10^{-14}$\\
  & $a_1=-2.3088\times 10^{-17}$\\
  & $a_2=7.3428\times 10^{-21}$\\
  & $a_3=-7.5474\times 10^{-25}$\\
  & $a_4=3.3468\times 10^{-29}$\\
  & $a_5=-5.528\times 10^{-34}$\\
42) $\mathrm{H_2}^++\mathrm{H}^-\longrightarrow \mathrm{H}+\mathrm{H_2}$ & $5\times10^{-12}T_{\rm g}^{-0.5}$ for $T_{\rm g} < 100$ & AAZN97 \\
43) $\mathrm{H_2}+\mathrm{e}^-\longrightarrow \mathrm{H}+\mathrm{H}^-$ &$3.67\times10^{-5}T_{\rm g}^{-2.28}e^{-47172/T_{\rm g}}$ & CCDL07 \\
44) $\mathrm{H_2}^++h\nu\longrightarrow \mathrm{H}+\mathrm{H}^+$ & &\\
thermal & $1.63\times10^7e^{-32400/T_{\rm r}} $ & GP98 \\
non-thermal: atom. recombin.&$\log k=\sum_{n=0}^5 a_n(\mathrm{log}T_{\mathrm r})^n$ &this work\\
&$a_0=-257.413$\\
&$a_1=294.406$\\
&$a_2=-93.7846$\\       
&$a_3=-13.1607$\\      
&$a_4=11.3683$\\      
&$a_5=-1.46734$\\
non-thermal: H$_2$ radiative cascade&$\log k=\sum_{n=0}^5 b_n(\mathrm{log} T_{\mathrm r})^n$ &this work\\
&$b_0=-1084.08$\\
&$b_1=1990.32$\\ 
&$b_2=-1447.42$\\  
&$b_3=503.994$\\ 
&$b_4=-83.6462$\\  
&$b_5=5.28898$\\ 
45) $\mathrm{H_2}+h\nu\longrightarrow \mathrm{H_2}^++\mathrm{e}^-$ & $3.06587\times 10^9 e^{-18948.1/T_{\rm r}}$ & C11 \\
46) $\mathrm{H_2^+}+h\nu\longrightarrow 2\mathrm{H}^++\mathrm{e}^-$ &$9\times10^1 T_{\rm r}^{1.48}e^{-335000/T_{\rm r}}$ & GP98 \\
47) $\mathrm{H_2}+h\nu\longrightarrow \mathrm{H_2}^*\longrightarrow2\mathrm{H}$ & $\ln k={17.555+7.2643\times 10^{-6}T_{\rm r}-1.4194\times 10^5T_{\rm r}^{-1}}$ & C11 \\
48) $\mathrm{H_2}+\mathrm{H} \rightarrow \mathrm{H}+\mathrm{H}+\mathrm{H}$ & 
$1.9535\times 10^{-10} T_{\rm g}^{-0.93267}e^{-49743/T_{\rm g}}$ & C11 \\
49) $\mathrm{D}+\mathrm{H_2}\longrightarrow \mathrm{HD}+\mathrm{H}$ &$1.69\times 10^{-16}e^{-4680 T_{\rm g}+198800/T_{\rm g}^2}$ for $T_{\rm g}>200$ & GP02 \\
  & $9\times 10^{-17}e^{-3876/T_{\rm g}}$ for $T_{\rm g}<200$ & GP98 \\
50) $\mathrm{D}^++\mathrm{H_2}\longrightarrow \mathrm{HD}+\mathrm{H}^+$ & $10^{-15}[0.417+0.846\log T_{\rm g}-0.137(\log T_{\rm g})^2]$ & GP02 \\
51) $\mathrm{HD}+\mathrm{H}\longrightarrow \mathrm{D}+\mathrm{H_2}$ &$5.25\times 10^{-17}e^{-4430/T_{\rm g}+173900/T_{\rm g}^2}$ for $T_{\rm g}>200$ & GP02 \\
  &$3.2\times 10^{-17}e^{-3624/T_{\rm g}}$ for $T_{\rm g}<200$ & GP98 \\
52) $\mathrm{HD}+\mathrm{H^+}\longrightarrow \mathrm{D^+}+\mathrm{H_2}$ & $1.1\times10^{-15}e^{-488/T_{\rm g}}$ & GP02 \\
53) $\mathrm{He}+\mathrm{H_2^+}\longrightarrow \mathrm{HeH^+}+\mathrm{H}$ &$3\times10^{-16}e^{-6717/T_{\rm g}}$ & GP98 \\
54) $\mathrm{HeH}^++\mathrm{H}\longrightarrow \mathrm{He}+\mathrm{H_2^+}$ & $4.3489\times10^{-16}T_{\rm g}^{0.110373} e^{-31.5396/T_{\rm g}}$ & BTGG11 \\
55) $\mathrm{H}^++\mathrm{H_2}\longrightarrow \mathrm{H_3^+}+h\nu$ &$10^{-18}$ & GP98\\
56) $\mathrm{H_3^+}+\mathrm{e}^-\longrightarrow \mathrm{H}+\mathrm{H_2}$ & $4.6\times10^{-12} T_{\rm g}^{-0.65}$ & GP98 \\
57) $\mathrm{H_2^+}+\mathrm{H}\longrightarrow \mathrm{H}+\mathrm{H}^++\mathrm{H}$&$\ln k=a_0+a_1T_{\rm g}+a_2T_{\rm g}^{-1}+a_3 T_{\rm g}^2$ & C11 \\
  & $a_0=-32.912$\\
  & $a_1=6.9498\times 10^{-5}$\\
  & $a_2=-3.3248\times 10^4$\\
  & $a_3=-4.08\times 10^{-9}$\\ 
58) $\mathrm{H_2}+\mathrm{H}^+\longrightarrow \mathrm{H}+\mathrm{H}+\mathrm{H}^+$&$\ln k=a_0+a_1T_{\rm g}+a_2T_{\rm g}^{-1}+a_3 T_{\rm g}^2$ & C11 \\
  & $a_0=-33.404$\\
  & $a_1=2.0148\times 10^{-4}$\\
  & $a_2=-5.2674\times 10^4$\\
  & $a_3=-1.0196\times 10^{-8}$\\ 
\hline
\end{tabular}
\end{minipage}
\end{table*}


\end{document}